\documentclass[prl,twocolumn,showpacs]{revtex4}     
\usepackage{epsfig,amsmath} 

\begin{document}

\title{Hall Coefficient of Dirac Fermions in Graphene under Charged Impurity Scatterings} 

\author{Xin-Zhong Yan$^{1,2}$ and C. S. Ting$^2$}
\affiliation{$^{1}$Institute of Physics, Chinese Academy of Sciences, P.O. Box 603, 
Beijing 100190, China\\
$^{2}$Texas Center for Superconductivity, University of Houston, Houston, Texas 77204, USA}
 
\date{\today}
 
\begin{abstract}
With a conserving formalism within the self-consistent Born approximation, we study the Hall conductivity of Dirac fermions in graphene under charged impurity scatterings. The calculated inverse Hall coefficient is compared with the experimental data. It is shown that the present calculations for the Hall coefficient and the electric conductivity are in good agreement with the experimental measurements.
\end{abstract}

\pacs{73.50.-h, 72.10.Bg, 81.05.Uw, 75.47.-m} 

\maketitle

Since the experiments of graphene were realized \cite{Novoselov,Geim,Zhang}, much effort has been devoted to studying the transport properties of the Dirac fermions. Many theoretical studies are based on the model of zero-range scatters in graphene \cite{Ziegler,Shon,Zheng,McCann,Khveshchenko,Aleiner,Peres,Ostrovsky}. However, it has been found that the charged impurities with screened Coulomb potentials \cite{Nomura,Hwang,Yan} are responsible for the observed carrier density dependence of the electric conductivity of graphene \cite{Geim}. Although the electric conductivity experiment has been successfully explained, so far there existed no satisfactory theories to fit the density dependence of inverse Hall coefficient as measured by another experiment in Ref. \onlinecite{Geim}.  

In this work, with a conserving formalism within the self-consistent Born approximation (SCBA), the Hall conductivity is calculated by using the diagrams generated from the current-current correlation function. We show that the experimentally measured electric conductivity and the inverse Hall coefficient can both be successfully explained in terms of the carrier scatterings off charged impurities.  

At low carrier concentration, the low energy excitations of electrons in graphene can be viewed as massless Dirac fermions \cite{Wallace,Ando,Castro,McCann} as being confirmed by recent experiments \cite{Geim,Zhang}. Using the Pauli matrices $\sigma$'s and $\tau$'s to coordinate the electrons in the two sublattices ($a$ and $b$) of the honeycomb lattice and two valleys (1 and 2) in the first Brillouin zone, respectively, and suppressing the spin indices for briefness, the Hamiltonian of the system is given by
\begin{equation}
H = \sum_{k}\psi^{\dagger}_{k}v\vec
 k\cdot\vec\sigma\tau_z\psi_{k}+\frac{1}{V}\sum_{kq}\psi^{\dagger}_{k-q}V_i(q)\psi_{k} \label{H}
\end{equation}
where $\psi^{\dagger}_{k}=(c^{\dagger}_{ka1},c^{\dagger}_{kb1},c^{\dagger}_{kb2},c^{\dagger}_{ka2})$ is the fermion operator, the momentum $k$ is measured from the center of each valley, $v$ ($\sim$ 5.86 eV\AA) is the velocity of electrons, $V$ is the volume of system, and $V_i(q)$ is the Thomas-Fermi-type charged impurity potential \cite{Yan}. Here, we neglect the intervalley scatterings in $V_i(q)$ for two reasons. First, for low electron doping, the intervalley scatterings are negligible small than the intravalley ones. Second, by doing so, our formulation of the problem given below will be much simplified.

\begin{figure} 
\centerline{\epsfig{file=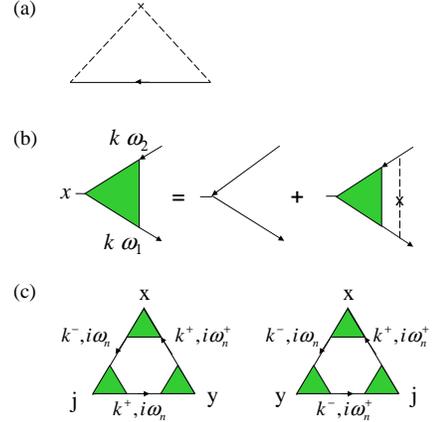,width=6.2 cm}}
\caption{(color online) (a) Self-consistent Born approximation for the self-energy. (b) Current vertex with impurity insertions. (c) Diagrams for calculating the Hall conductivity $\sigma_{xy}$. The vertex $j$ is associated with the vector potential $A_j$. $k^{\pm} = k \pm q/2$. $\omega^+_n =\omega_n+\Omega_m$ with $\Omega_m$ and $\omega_n$ respectively the bosonic and fermionic Matsubara frequencies.} 
\end{figure} 

Under the SCBA [see Fig. 1(a)] \cite{Fradkin,Lee1}, the Green function $G(k,\omega)=[\omega+\mu-v\vec
 k\cdot\vec\sigma\tau_z-\Sigma(k,\omega)]^{-1}$ $\equiv g_0(k,\omega)+ g_c(k,\omega)\hat k\cdot\vec\sigma\tau_z$ and the self-energy $\Sigma(k,\omega)$ of the single particles are determined by coupled integral equations \cite{Yan}. Correspondingly, the current vertex correction is the ladder diagrams in Fig. 1(b). The current vertex $v\Gamma_x(k,\omega_1,\omega_2)$ is expanded as
\begin{equation}
\Gamma_x(k,\omega_1,\omega_2)=\sum_{j=0}^3y_j(k,\omega_1,\omega_2)A^x_j(\hat k) \label{vt}
\end{equation}
where $A^x_0(\hat k)=\tau_z\sigma_x$, $A^x_1(\hat k)=\sigma_x\vec\sigma\cdot\hat k$, $A^x_2(\hat k)=\vec\sigma\cdot\hat k\sigma_x$, $A^x_3(\hat k)=\tau_z\vec\sigma\cdot\hat k\sigma_x\vec\sigma\cdot\hat k$, and $y_j(k,\omega_1,\omega_2)$ are determined by four-coupled integral equations \cite{Yan}. 

We start to calculate the Hall coefficient of the Dirac-fermion system in graphene. Consider that the system is acted with an weak external magnetic field $B$ perpendicular to the graphene plane. The vector potential $\vec A(q)$ is related to $B$ via $\vec B = i\vec q\times\vec A(q)$ where $\vec q$ is a wave vector. The Hall conductivity $\sigma_{xy}$ is defined as the ratio between the electric current density along $x$ direction and the electric field applied in $y$ direction. By the standard linear response theory, $\sigma_{xy}$ is obtained as $\sigma_{xy} = -\lim_{\Omega\to 0}{\rm Im}\Phi_{xy}(\Omega+i0^+)/\Omega$
where $\Phi_{xy}(\Omega)$ is the current correlation function. For the Matsubara frequency $\Omega_m = 2\pi mT$ with $m$ as a integer, $\Phi_{xy}(i\Omega_m)$ is given by \cite{Fukuyama}
\begin{widetext}
\begin{eqnarray}
\Phi_{xy}(i\Omega_m) = -\frac{1}{V}\sum_j\int_{0}^{\beta}d\tau\int_{0}^{\beta}d\tau'e^{i\Omega_m\tau}\langle T_{\tau}J_x(q,\tau)J_y(0,0)J_j(-q,\tau')\rangle A_j(q)
\end{eqnarray}
where $T_{\tau}$ is the $\tau$ ordering operator, $J_x(q,\tau)$ is the current operator, $\beta = T^{-1}$,  $\langle\cdots\rangle$ means the statistic average, and the use of units in which $\hbar = c = k_B = 1$ has been made. Within SCBA, the function $\Phi_{xy}(i\Omega_m)$ is calculated according to Fig. 1(c). Writing it explicitly, we have
\begin{eqnarray}
\Phi_{xy}(i\Omega_m) &=& \frac{2v^3e^3}{V\beta}\sum_{knj}{\rm Tr}\{[\Gamma_x(k^-,k^+,i\omega_n,i\omega_n^+)G(k^+,i\omega_n^+)\Gamma_y(k^+,k^+,i\omega_n^+,i\omega_n)
+\Gamma_y(k^-,k^-,i\omega_n,i\omega_n^-)\nonumber\\
&&\times G(k^-,i\omega_n^-)\Gamma_x(k^-,k^+,i\omega_n^-,i\omega_n)]
 G(k^+,i\omega_n)\Gamma_j(k^+,k^-,i\omega_n,i\omega_n)G(k^-,i\omega_n)\}A_j(q) \label{phi}
\end{eqnarray}
where the factor 2 stems from the spin degeneracy, $k^{\pm} = k\pm q/2$, and $\omega_n^{\pm} = \omega_n\pm \Omega_m$. The vertex given by Eq. (\ref{vt}) corresponds to $\Gamma_x(k,k,\omega_1,\omega_2)\equiv\Gamma_x(k,\omega_1,\omega_2)$. $\Gamma_{\mu}(k^-,k^+,\omega_1,\omega_2)$ satisfies the $4\times 4$ matrix equation
\begin{eqnarray}
\Gamma_{\mu}(k^-,k^+,\omega_1,\omega_2) = \tau_3\sigma_{\mu}+\frac{1}{V}\sum_{k_1}n_iv^2_0(k-k_1) G(k^-_1,\omega_1)\Gamma_{\mu}(k^-_1,k^+_1,\omega_1,\omega_2)G(k^+_1,\omega_2). \label{vt2} 
\end{eqnarray} 
\end{widetext}
$\Gamma_{\mu}(k^+,k^-,\omega_1,\omega_2)$ is obtained by exchanging - and + in Eq. (\ref{vt2}). The analytical continuation $i\Omega_m \to \Omega+i0^+$ for $\Phi_{xy}(i\Omega_m)$ can be manipulated according to the text book \cite{Mahan}. For the weak magnetic field $B$, we expand the right hand side of Eq. (\ref{phi}) to the linear terms of $q$ (giving rise to the linear terms of $B$). To do so, we need to note the facts listed below. (1) $\Gamma_{\mu}(k^-,k^+,\omega_1,\omega_2)$ is expanded as $\Gamma_{\mu}(k,\omega_1,\omega_2)+ \gamma_{\mu}(k,q,\omega_1,\omega_2)$ with 
\begin{widetext}
\begin{eqnarray}
\gamma_{\mu}(k,q,\omega_1,\omega_2)&=&\frac{1}{V}\sum_{k'}n_iv^2_0(k-k') G(k',\omega_1)\gamma_{\mu}(k',q,\omega_1,\omega_2)G(k',\omega_2) \nonumber\\
&&-\frac{1}{V}\sum_{k'}n_iv^2_0(k-k')[\nabla G(k',\omega_1)\Gamma_{\mu}(k',\omega_1,\omega_2)G(k',\omega_2)
-G(k',\omega_1)\Gamma_{\mu}(k',\omega_1,\omega_2)\nabla G(k',\omega_2)]\cdot \vec q/2  \nonumber\\
\label{dw}
\end{eqnarray} 
where $\nabla$ means the gradient with respect to $k'$. (2) From the identity
\begin{eqnarray}
\frac{1}{V}\sum_{kk'}n_i[(\frac{\partial}{\partial k_j}+\frac{\partial}{\partial k'_j}) v^2_0(k-k')] {\rm Tr}[G(k,\omega_1)\gamma_{\mu}(k,q,\omega_1,\omega_2)G(k,\omega_2)G(k',\omega_2)\Gamma_{\nu}(k',\omega_2,\omega_1)G(k',\omega_1)] = 0, \nonumber
\end{eqnarray} 
rewriting the left hand side by performing the integral by part and using the equations for $\Gamma_{\nu}(k,\omega_1,\omega_2)$ and $\gamma_{\mu}(k,q,\omega_1,\omega_2)$, we obtain
\begin{eqnarray}
\sum_k{\rm Tr}\{\gamma_{\mu}(k,q,\omega_1,\omega_2)[\frac{\partial}{\partial k_j}G(k,\omega_2)\Gamma_{\nu}(k,\omega_2,\omega_1)G(k,\omega_1)+ G(k,\omega_2)\Gamma_{\nu}(k,\omega_2,\omega_1)\frac{\partial}{\partial k_j}G(k,\omega_1)]\}\nonumber\\
= -\frac{\vec q}{2}\cdot\sum_{k}{\rm Tr}\{[\nabla G(k,\omega_1)\Gamma_{\mu}(k,\omega_1,\omega_2)G(k,\omega_2)  - G(k,\omega_1)\Gamma_{\mu}(k,\omega_1,\omega_2)\nabla G(k,\omega_2)]\frac{\partial}{\partial k_j} \Gamma_{\nu}(k,\omega_2,\omega_1)\}.
\end{eqnarray} 
(3) Using Ward identity, $vG(k,\omega)\Gamma_{\alpha}(k,\omega,\omega)G(k,\omega) = \partial G(k,\omega)/\partial k_{\alpha}$, we have
\begin{eqnarray}
vG(k^+,\omega)\Gamma_{\alpha}(k^+,k^-,\omega,\omega)G(k^-,\omega) = \frac{\partial}{\partial k_{\alpha}} G(k,\omega)+\frac{\vec q}{2}\cdot\sum_jA^{\alpha}_j(\hat k)[a_j(k,\omega)\hat k+i\sigma_z b_j(k,\omega)\hat \phi], \label{we}
\end{eqnarray} 
where $\phi$ is the angle of $k$ and $\hat \phi$ is the unit vector in $\phi$ direction. This expansion can be further simplified. If we consider the left hand side as a functional of the matrices $A^{\alpha}_j(\hat k)$, then its transpose should be a functional of matrices $\tilde A^{\alpha}_j(\hat k)$ defined as
$\tilde A^{\alpha}_0(\hat k)=\tau_z\sigma^t_{\alpha}$,
$\tilde A^{\alpha}_1(\hat k)=\sigma^t_{\alpha}\vec\sigma^t\cdot\hat k$, 
$\tilde A^{\alpha}_2(\hat k)=\vec\sigma^t\cdot\hat k\sigma^t_{\alpha}$, and
$\tilde A^{\alpha}_3(\hat k)=\tau_z\vec\sigma^t\cdot\hat k\sigma^t_{\alpha}\vec\sigma^t\cdot\hat k$ where $\sigma^t$ means the transpose of the Pauli matrix. By comparing the transpose of Eq. (\ref{we}) and the expansion of $vG^t(k^-,\omega)\Gamma^t_{\alpha}(k^+,k^-,\omega,\omega)G^t(k^+,\omega)$ in terms of $\tilde A^{\alpha}_j(\hat k)$, we find $a_1 = -a_2 \equiv a$, $b_1 = b_2 \equiv b$, and all other $a$'s and $b$'s vanish. Using $\sigma_{\alpha}\sigma\cdot\hat k-\sigma\cdot\hat k\sigma_{\alpha}=-2i\sigma_z\hat\alpha\cdot\hat\phi$, and $\sigma_{\alpha}\sigma\cdot\hat k+\sigma\cdot\hat k\sigma_{\alpha}=2\hat\alpha\cdot\hat k$, we obtain 
\begin{eqnarray}
vG(k^+,\omega)\Gamma_{\alpha}(k^+,k^-,\omega,\omega)G(k^-,\omega) = \frac{\partial}{\partial k_{\alpha}} G(k,\omega)+i\sigma_z\hat\alpha\cdot[b(k,\omega)\hat k\hat\phi-a(k,\omega)\hat\phi\hat k]\cdot\vec q. \label{we1}
\end{eqnarray} 
This result can also be obtained by solving Eq. (\ref{dw}). Since the final result depends on $a(k,\omega)+b(k,\omega)\equiv c(k,\omega)$, we here present only the equations for determining $c(k,\omega)\equiv z(k,\omega)+[g^2_0(k,\omega)-g^2_c(k,\omega)]X(k,\omega)$: 
\begin{eqnarray}
z(k,\omega) &=& [g'_0(k,\omega)g_c(k,\omega)-g_0(k,\omega)g'_c(k,\omega)][y_0(k,\omega,\omega)-y_3(k,\omega,\omega)] \nonumber\\
&&~~-g_c(k,\omega)\{g_0(k,\omega)[y_0(k,\omega,\omega)+y_3(k,\omega,\omega)]+2g_c(k,\omega)y_1(k,\omega,\omega)\}/k ,\\
X(k,\omega) &=& \frac{1}{V}\sum_{k'}n_iv^2_0(k-k')\{z(k',\omega)+[g^2_0(k',\omega)-g^2_c(k',\omega)]X(k',\omega)\}, 
\end{eqnarray} 
where $g'(k,\omega)=\partial g(k,\omega)/\partial k$. Using above results, we obtain for the Hall conductivity
\begin{eqnarray}
\sigma_{xy} &=& \frac{Bv^2e^3}{V}\sum_k\int_{-\infty}^{\infty}\frac{d\omega}{2\pi}[-\frac{\partial F(\omega)}{\partial\omega}]{\rm Im Tr}\{
\Gamma_x^{-+}[G^+(\frac{\partial\Gamma_y^{+-}}{\partial k_x}\frac{\partial G^-}{\partial k_y}
-\frac{\partial\Gamma_y^{+-}}{\partial k_y}\frac{\partial G^-}{\partial k_x}) \nonumber\\
&& +(\frac{\partial G^+}{\partial k_x}\frac{\partial\Gamma_y^{+-}}{\partial k_y}
-\frac{\partial G^+}{\partial k_y}\frac{\partial\Gamma_y^{+-}}{\partial k_x})G^-
-2ic(k,\omega^-)G^+\Gamma_y^{+-}\sigma_z]\} \label{hcd}
\end{eqnarray} 
\end{widetext}
where $\Gamma_x^{-+} = \Gamma_x(k,\omega^-,\omega^+)$, $\Gamma_y^{+-} = \Gamma_y(k,\omega^+,\omega^-)$, and $G^{\pm} = G(k,\omega^{\pm})$. Some terms such as $\Gamma_x^{-+}(\partial G^+/\partial k_x\Gamma_y^{+-}\partial G^-/\partial k_y-\partial  G^+/\partial k_y\Gamma_y^{+-}\partial G^-/\partial k_x)$ and those containing the same frequency-$\omega^+$ arguments are not included because they happen to be zero under the operation ${\rm Im Tr}$ in the right hand side of Eq. (\ref{hcd}). 
 
\begin{figure} 
\centerline{\epsfig{file=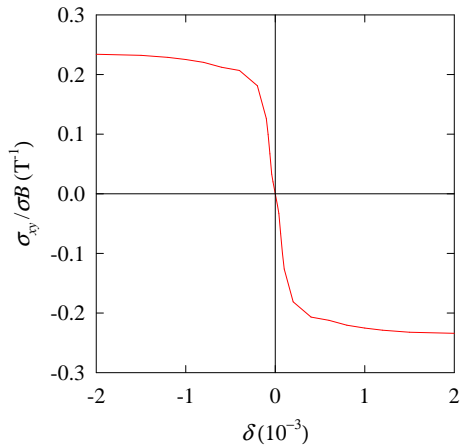,width=6.4 cm}}
\caption{(color online) Hall conductivity $\sigma_{xy}$ normalized by $\sigma B$ as function of the electron doping concentration $\delta$. $\sigma$ is the electric conductivity, and $B$ is the magnetic field in unit of Tesla.} 
\end{figure} 

The numerical result for the Hall conductivity $\sigma_{xy}$ as a function of carrier concentration $\delta$ (doped carrier per carbon atom) is shown in Fig. 2. Here the impurity density $n_i = 1.15\times 10^{-3}a^{-2}$ (with $a$ as the lattice constant of graphene) is chosen as the same as in our previous work \cite{Yan}. In the normalization factor $1/\sigma B$ given in Fig. 2, the magnetic field B is a constant and $\sigma$ is the electric conductivity (see Fig. 3). In the limit $\delta \to 0$, the normalization denominator is a constant because of the minimum conductivity. While at large carrier concentration, it is linear in $\delta$ because of $\sigma \propto \delta$. Within a very narrow range of $\delta$ around 0, $\sigma_{xy}$ varies dramatically and vanishes at $\delta = 0$. Beyond this regime, the saturation behavior of the curve implies that $\sigma_{xy} \propto \delta$. The present result differs quantitatively from Ref. \onlinecite{Zheng} using a different approach. Also, it is qualitative different from Ref. \onlinecite{Nakamura} in which a constant scattering rate was phenomenologically introduced and the current correlation was treated without vertex correction.

For calculating the Hall coefficient, we need the result of the electric conductivity $\sigma$ obtained with the same scattering parameters as for $\sigma_{xy}$. The result for $\sigma$ as a function of $\delta$ is shown in Fig. 3 and compared with experimental data \cite{Geim}. The minimum conductivity by the present calculation is about 3.5 $e^2/h$ close to the data 4 $e^2/h$, which is a consequence of the coherence between the upper and lower band states. Our numerical calculation shows that the contribution from intervalley scatterings is negligible small. 
 
\begin{figure} 
\centerline{\epsfig{file=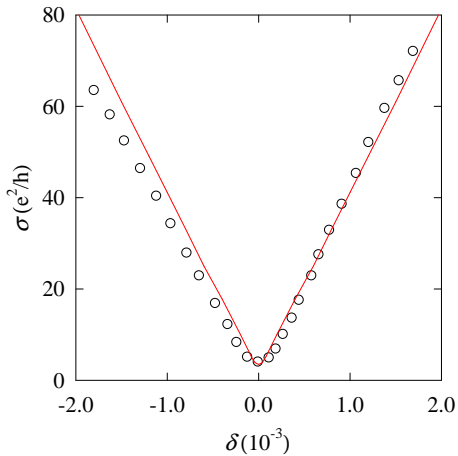,width=6.4 cm}}
\caption{(color online) Electric conductivity $\sigma$ as function of the electron doping concentration $\delta$. The present calculation (solid line) is compared with the experimental data in Ref. \onlinecite{Geim} (symbols).} 
\end{figure} 

In Fig. 4, we exhibit the theoretical result (solid line) for the inverse Hall coefficient defined as $R^{-1} = B\sigma^2/\sigma_{xy}$ and compare it with the experimental data \cite{Geim} (symbols). Clearly, the present calculation fits the experimental measurement of the inverse Hall coefficient as well as the electric conductivity very well. With comparing to the classical prediction $R^{-1} = -nec$ (with $n$ as the doped electron density), both the present calculation and the experiment data for $R^{-1}$ diverge at $\delta = 0$. The divergence of $R^{-1}$ stems from the vanishing of $\sigma_{xy}$ at $\delta =0$ while the conductivity $\sigma$ remains finite. The classical theory is based on the concept of the drift velocity. At the zero carrier concentration if the conductivity remains finite, the drift velocity has to become infinitively large, which implies an infinitely large Lorentz force acting on an electron. Therefore, the classical theory is not applicable near $\delta = 0$. On the other hand, at large carrier concentration, the present calculation reproduces the classical theory, and both of them are in agreement with the experimental result.
 
\begin{figure}
\centerline{\epsfig{file=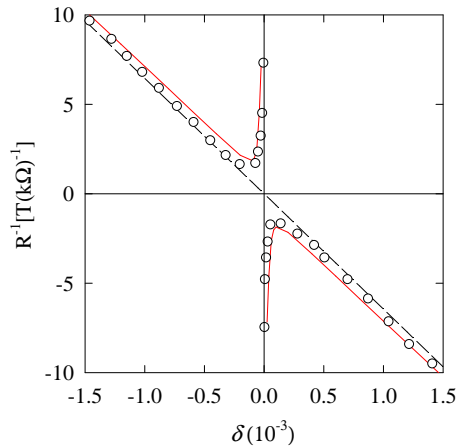,width=6.4 cm}} 
\caption{(color online) The inverse Hall coefficient $R^{-1}$ (in unit of $10^{-3}$Tesla/Ohm) as function of the electron doping concentration $\delta$. The present calculation (solid line) is compared with the experimental data in Ref. \onlinecite{Geim} (symbols). The dashed line represents the classic theory $R^{-1} = -nec$. } 
\end{figure} 
 
In summary, on the basis of self-consistent Born approximation, we have calculated the Hall coefficient of the Dirac fermions under the charged impurity scatterings in graphene. The anomalous in the inverse Hall coefficient at zero carrier concentration stems from the vanishing of the Hall conductivity and meanwhile the minimum remained in the electric conductivity. The present results for the inverse Hall coefficient and the electric conductivity are in very good agreement with the experimental measurements. 

This work was supported by a grant from the Robert A. Welch Foundation under No. E-1146, the TCSUH, the National Basic Research 973 Program of China under grant No. 2005CB623602, and NSFC under grant No. 10774171 and No. 10834011.

\end{document}